## Linking Topological Medium Range Order and Density-Wave Coherence to Metallic Glass Ductility

Minhazul Islam[1], Muchen Wang[2], Andrea Fantin[3], Geun Hee Yoo[4], Ji Young Kim[4], Hamdan Ashfaq[5], Eun Soo Park[4], Robert Maass[3], Yunzhi Wang[1], Yue Fan[2], Hee-Suk Chung[6], Sang-Chul Lee[6,†], Jinwoo Hwang[1,†]

1. Department of Materials Science and Engineering, The Ohio State University, Columbus OH 43212, USA.
2. Department of Mechanical Engineering, University of Michigan, Ann Arbor, MI 48104, USA.
3. Federal Institute of Materials Research and Testing, Unter den Eichen 87, 12205 Berlin, Germany.
4. Department of Materials Science and Engineering, Research Institute of Advanced Materials & Institute of Engineering Research, Seoul National University, Seoul 08826, Republic of Korea.
5. Department of Computer Science and Data Analytics, Denison University, Granville, OH 43023, USA.
6. Research Center for Materials Analysis, Korea Basic Science Institute, Daejeon, Republic of Korea.

([†] *Corresponding Authors*)

**Abstract**

Metallic glasses exhibit pronounced composition-dependent mechanical properties, yet identifying the structural origins of these changes remains difficult. An outstanding issue is how to correlate different forms of medium range order (MRO), defined at different length scales and accessed through different experimental observables, and to understand their connection to properties. We combine synchrotron reduced density function, $G(r)$, with machine-learning-assisted four dimensional scanning transmission electron microscopy (4D-STEM) to examine geometric and topological MRO in five Zr-based metallic glasses. Density wave analysis of $G(r)$ provides a geometric MRO descriptor through the coherence length of atomic density correlations, while 4D-STEM reveals nanoscale topological and chemical MRO motifs, including crystal-like local ordering with distinct rotational symmetries. We show that the $G(r)$-derived coherence length captures an important baseline tendency for cooperative ductile relaxation, while composition-dependent topological MRO provides an additional structural contribution that helps explain the observed mechanical responses. In particular, increasing Zr content suppresses Cu-rich FCC-like MRO, increases the cooperative coherence volume, and promotes fine Zr-rich HCP-like local order associated with enhanced ductility. These results identify the combined structural influence of geometric and topological MRO as a key basis for composition-dependent mechanical response in metallic glasses.

## 1. Introduction

Understanding the atomic structure of metallic glasses (MGs) and its connection to composition-dependent properties has long been a central challenge in glass science. In many MG systems, even modest compositional changes can produce pronounced variations in mechanical response [1]. Zr-based MGs provide a particularly striking example, where changes in composition [2, 3] can shift the plasticity limit from essentially zero to a superplastic regime. This strong compositional sensitivity underscores the need to identify the structural features that govern mechanical behavior, particularly how atomic arrangements at the short range order (SRO) and medium range order (MRO) scales contribute to these property changes.

Computational studies have proposed various atomic-scale models representing a hierarchy of length scales and motifs both at SRO and MRO scales, *e.g.* [4-7]. Experimental validation of such structures and their evolution in real MGs, however, has remained a greater challenge. Large-area diffraction techniques combined with pair distribution function analyses [8] have provided important insights into the atomic structure of glass-forming liquids and a variety of MG compositions, *e.g.* [9]. More recently, density wave analysis [10-13] has been applied to the reduced density function, $G(r)$, to extract the structural coherence length, which represents the MRO length scale over which density fluctuations extend. In this framework, the MRO extracted from $G(r)$ can be considered “geometric MRO”, because it describes the real-space geometry of atomic density correlations, rather than the topology of specific atomic motifs [10]. The coherence length has been shown to correlate with properties of many metallic glasses, including the fragility index, which has also been found to have an approximate relationship with the overall plasticity limit of the material.

Meanwhile, recent four dimensional scanning transmission electron microscopy (4D-STEM) experiments have revealed details of MG structure in a spatially resolved manner at the nanoscale [14-21]. Evolved from analyses based on fluctuation electron microscopy [22-27], these 4D-STEM approaches provide more detailed insight into the nanostructure and offer new perspectives on structure-property relationships. Analysis methods such as angular correlation functions [15, 16, 28, 29], power spectrum analysis [14, 21], and Pearson correlation analysis [17] have revealed the structural symmetry, size, and distribution of nanoscale volumes that generate relatively strong diffraction intensity. These nanovolumes can therefore be viewed as “topological MRO”, because they reflect the local arrangement, symmetry, and connectivity of atoms beyond the nearest-neighbor shell. These studies have shown that nanodiffraction signals from MGs often exhibit distinct structural symmetries consistent with crystal-like ordering, which had previously been observed only in limited computational studies *e.g.* [30], and may contain rich details of nanoscale heterogeneity in MGs. However, it should also be noted that the current 4D-STEM analysis methods may have detection limits, as they can be less sensitive to weaker signals arising from more subtle atomic arrangements. Another challenge is that 4D-STEM generates a massive amount of data, often at least a few hundred thousand nanodiffraction patterns per sample, making it inherently difficult to identify detailed features in each pattern.

Topological MRO and the aforementioned geometric MRO, each associated with different characterization approaches, represent two distinct definitions of MRO structure, making their relationship an important question. Studies have shown that the glass transition freezes geometric MRO at a larger scale, while topological MRO represents shorter-scale structure that remains more “liquid-like” even after the glass transition, with its formation governed by short-range chemical bonding characteristics [10]. Based on this framework, an important question is how topological MRO contributes to the coherence length determined from $G(r)$, and how this relationship connects to the mechanical properties of MGs. Although the coherence length extracted from $G(r)$ is commonly interpreted as a geometric MRO descriptor, characteristics of topological MRO, including structural symmetry, size, and spatial distribution, may also influence the measured density correlations and provide a distinct structural contribution to mechanical response.

Using 4D-STEM analysis and density wave analysis of *G*(*r*) from synchrotron diffraction, we report a comprehensive study of nanoscale MRO structure in five Zr-based MG compositions, spanning the binary $Zr_{50}Cu_{50}$, eutectic ternary $Zr_{50}Cu_{40}Al_{10}$, hypoeutectic $Zr_{60}Cu_{30}Al_{10}$ and $Zr_{65}Cu_{25}Al_{10}$, and quinary $Zr_{52.5}Cu_{17.9}Ni_{14.6}Al_{10}Ti_{5}$ (Vit-105), which exhibits a wide range of properties. A key technical novelty of this work is the first use of machine learning (ML) to identify rotational symmetries directly from individual amorphous nanodiffraction patterns from 4D-STEM. This represents a technically demanding classification problem because the rotational symmetry signatures in amorphous nanodiffraction patterns are typically weak, diffuse, and partially obscured by disordered scattering, unlike the sharp Bragg reflections typically used for symmetry identification in crystalline diffraction (e.g. [31]). The successful classification of these subtle rotational features therefore represents an important methodological advance, enabling spatially resolved mapping and statistical analysis of nanoscale MRO motifs that would be difficult to obtain through averaged angular correlation analysis alone.

The combined use of 4D-STEM analysis and density-wave analysis of synchrotron *G*(*r*) reveals how geometric and topological MRO are related and how their combined influence affects mechanical response in Zr-based MGs. We first show that the density-wave coherence length extracted from *G*(*r*) establishes a baseline tendency for ductile relaxation. We then show that 4D-STEM probes a distinct but related structural level, revealing nanoscale topological and chemical MRO motifs that are not resolved by the averaged density correlation alone. By comparing compositions with different chemical complexity and Zr/Cu ratios, we show that similar geometric coherence lengths do not necessarily lead to similar ductility, and that local motif topology modifies how geometric coherence is expressed mechanically. Specifically, large Cu-rich FCC-like MRO can reduce structural coherence and promote brittle behavior, whereas fine Zr-rich HCP-like MRO can diversify the nanoscale structure, frustrate strain localization, and facilitate repeated shear-band relaxation. Overall, this work shows that metallic-glass ductility reflects the combined influence of geometric MRO, which sets the cooperative length scale for relaxation, and topological MRO, whose size, chemistry, and spatial distribution provide an additional structural basis for understanding the observed differences in plasticity.

## 2. Methodology

### 2.1. Sample preparation

Metallic glass ingots with compositions $Zr_{50}Cu_{50}$, $Zr_{50}Cu_{40}Al_{10}$, $Zr_{60}Cu_{30}Al_{10}$, $Zr_{65}Cu_{25}Al_{10}$ and $Zr_{52.5}Cu_{17.9}Ni_{14.6}Al_{10}Ti_{5}$ (Vit-105) were prepared by arc melting, followed by suction casting into cylindrical rods with a diameter of 2 mm. TEM samples were prepared using a focused ion beam (FIB, Thermo Fisher Helios) as described in our previous reports [15, 16]. Particular emphasis was given to achieving thin and uniform TEM foil thickness as well as surface cleanliness, which was accomplished using low-energy ion milling (Fischione Nanomill) after the FIB process, as the thickness and surface quality can critically affect the quality of 4D-STEM data. The electron transmittance [32] of the FIB foils was within the range of $75 \pm 2$ %, corresponding to about 25 nm in real thickness.

### 2.2. Synchrotron reduced density function and coherence length

$G(r)$'s and structure factors, $S(q)$'s, from $Zr_{50}Cu_{50}$, $Zr_{50}Cu_{40}Al_{10}$, and $Zr_{65}Cu_{25}Al_{10}$ MGs were acquired from synchrotron diffraction data from Advanced Photon Source. $G(r)$'s from $Zr_{60}Cu_{30}Al_{10}$ and Vit-105 MG were acquired from synchrotron diffraction data from Technical University of Munich. The structural coherence length is determined from the $G(r)$'s. It represents the distance over which atomic density correlations persist in a volume-averaged sense. Following the coherence model for metallic glasses, the medium-range contribution to *G*(*r*) can be expressed as a damped oscillatory term [10],

$$G(r) = 4\pi r\rho_0\{g(r) - 1\} = 4\pi r\rho_0 \left\{A_{MRO}\,\frac{a}{r}\sin(Q_{MRO}r + \delta_{MRO})\,e^{-r/\zeta}\right\}, \qquad (1)$$

where *g*(*r*) is the pair distribution function, and $\zeta$ is the structural coherence length that characterizes the spatial persistence of medium-range density correlations beyond the short-range cutoff distance. $A_{MRO}$, $a$, $Q_{MRO}$, and $\delta_{MRO}$ are fitting parameters that approximately represent different aspects of MRO. In particular, $a$ is the average nearest-neighbor distance determined from the first peak position of *G*(*r*). $\rho_0$ is the average density of the material. The corresponding coherence volume is $(\zeta/a)^3$.

**2.3. 4D-STEM experiment**

4D-STEM experiments were conducted using a Thermo Fisher Themis STEM operated at 300 kV. A probe size of 1 nm was used, achieved by selecting a C2 aperture of 10 $\mu$m and a convergence semi-angle of 2.23 mrad, with a beam current of 50 pA. Nanodiffraction patterns were recorded using a Thermo Fisher EMPAD detector with 32-bit depth. Each dataset was acquired from sample regions using oversampled probe positions, with a probe dwell time of 1 ms. A typical scan covered a 256 by 256 grid of probe locations, resulting in 65,536 diffraction patterns per dataset. This scan protocol was repeated over multiple regions for each sample. The sample thickness was controlled to ensure about 75% electron transmittance to minimize plural scattering effects, while remaining below 80% to avoid excessive signal contributions from surface oxidation. New 4D-STEM data were acquired from $Zr_{60}Cu_{30}Al_{10}$ and Vit-105 (Vitreloy 105 [33]) in this work. All data sets, including the previously acquired 4D-STEM data from $Zr_{50}Cu_{50}$, $Zr_{50}Cu_{40}Al_{10}$, and $Zr_{65}Cu_{25}Al_{10}$ [16], were analyzed using both angular correlation analysis and the ML method as described below. Angular correlation measures the autocorrelation between the pixel intensities in each nanodiffraction pattern as a function of the azimuthal angle (0 to 2π) [14, 15]. The resulting patterns were then averaged to reveal dominant structural symmetries in each composition.

### 2.4. Machine learning analysis of 4D-STEM data

Angular correlation analyses show that 4D-STEM nanodiffraction patterns contain symmetry features that reflect local atomic ordering in metallic glasses. Identifying these symmetries enables characterization of their size and spatial distribution, but the large number of patterns in typical datasets makes manual analysis impractical. To address this, we employed a supervised ML classification framework operating directly on raw diffraction patterns, allowing categorization of symmetry features without relying on crystallographic models.

From datasets containing up to several hundred thousand patterns per composition, 2,000-3,000 patterns were initially labeled by visual inspection based on rotational symmetries such as 2-fold, 4-fold, and 6-fold, along with an "other" category for diffuse or unidentifiable patterns. Higher-order symmetries were also identified during iterative training. The model is based on the Swin Transformer V2 (Large) architecture, pretrained on large-scale image datasets. It applies hierarchical feature learning through multi-stage self-attention, capturing both local features and global symmetry information. Diffraction patterns were resampled from 128 × 128 to 384 × 384 pixels and fine-tuned using pretrained weights. Data augmentation included intensity scaling, Poisson and Gaussian noise, Gaussian blur, rotation, and horizontal flipping. A double circular mask was applied to remove the central beam and background, improving signal-to-noise while preserving symmetry.

Training was performed in PyTorch using the TIMM framework, with CrossEntropyLoss and AdamW optimizer. The learning rate followed a OneCycleLR schedule from $5 \times 10^{-5}$ to $5 \times 10^{-4}$, with weight decay of 0.05. Data were split 80:20 for training and validation. The model outputs class probabilities for each pattern, and the top-1 prediction was used to construct real-space symmetry maps and correlate symmetry with local structure and mechanical behavior.

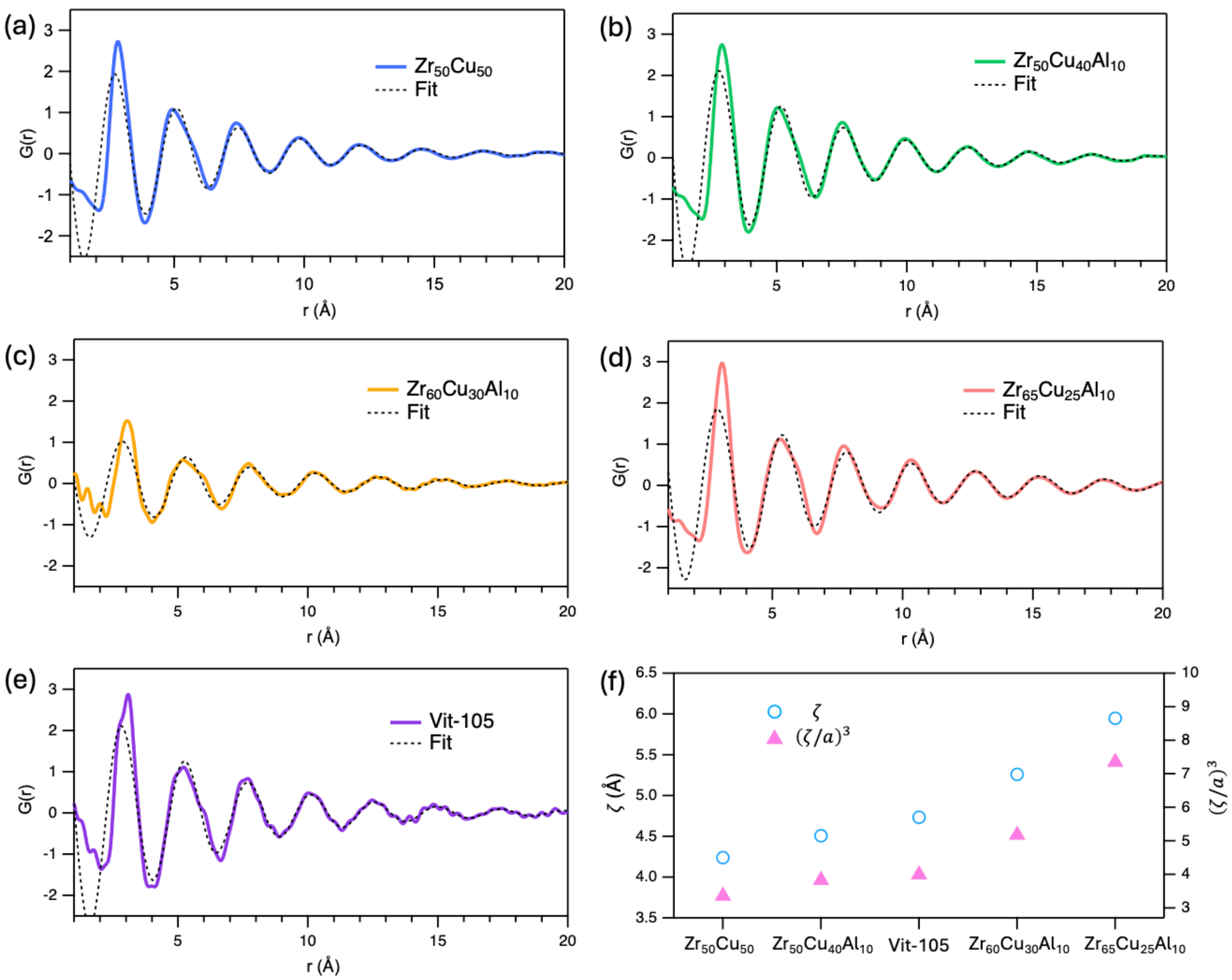

**Figure 1**. (a to e) $G(r)$ and fit for coherence length extraction using the density wave framework. (f) Structural coherence length, $\zeta$, and volume, $(\zeta/a)^3$, extracted from the fitting functions.

## 3. Results and Discussion

### 3.1. Coherence length from reduced density function

Figure 1a to 1e shows the experimental $G(r)$ from all five compositions, along with the density wave fitting functions to extract the structural coherence length, $\zeta$, a parameter representing the degree of geometric MRO. The fit was optimized at larger $r$, as the density wave formalism describes the exponential decay of medium-range density correlations rather than short-range atomic packing. As a result, deviations at the first peak are expected, since these features are dominated by short-range chemical order not captured by the model [10]. The extracted coherence lengths, $\zeta$ , are summarized in Fig. 1f (open circles). A clear variation in $\zeta$ is observed across the five compositions. $Zr_{50}Cu_{50}$ exhibits the shortest $\zeta$ of $4.23 \pm 0.04$ Å, where the uncertainty arises from the fitting error. $\zeta$ shows an increasing trend with Zr content, with $Zr_{65}Cu_{25}Al_{10}$ showing the highest $\zeta$ of $5.94 \pm 0.06$ Å. A similar trend is shown for the normalized coherence volume, $(\zeta/a)^3$ (closed triangle). The coherence length and normalized coherence volume measured in these alloys correlate with their ductility, as will be discussed in detail in Section 3.5.

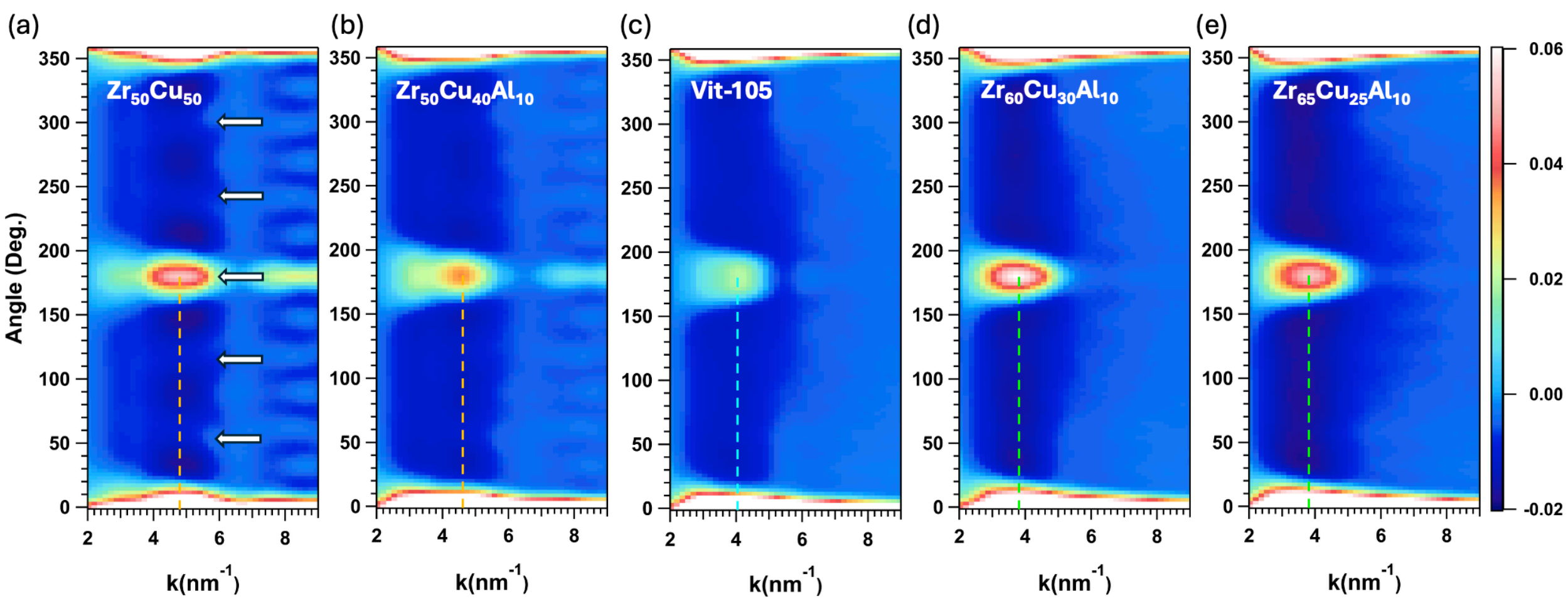

**Figure 2.** Averaged angular correlation from (a) $Zr_{50}Cu_{50}$, (b) $Zr_{50}Cu_{40}Al_{10}$, (c) Vit-105, (d) $Zr_{60}Cu_{30}Al_{10}$, and (e) $Zr_{65}Cu_{25}Al_{10}$.

### 3.2. 4D-STEM angular correlation analysis

The angular correlation functions for all five compositions in Fig. 2 show how the characteristics of both topological and geometric MRO change across different compositions. Among them, the data from $Zr_{60}Cu_{30}Al_{10}$ and Vit-105 were newly acquired in the present study, and the data from $Zr_{50}Cu_{50}$, $Zr_{50}Cu_{40}Al_{10}$, and $Zr_{65}Cu_{25}Al_{10}$ are the same as those included in our previous work [16]. $Zr_{50}Cu_{50}$ and $Zr_{50}Cu_{40}Al_{10}$ (Fig. 2a and 2b) exhibit a broad 2-fold peak (180°) at ~ 4.8 $nm^{-1}$ (yellow dashed lines), as well as notable 6-fold symmetry (white arrows in Fig. 2a), which is consistent with the presence of a high degree of Cu-rich, crystal-like topological MRO [16]. The fact that $Zr_{50}Cu_{50}$ has higher peak intensity than $Zr_{50}Cu_{40}Al_{10}$ indicates that the two alloys have different overall characteristics of Cu-rich crystal-like MRO. More details of this trend will be revealed in ML-assisted 4D-STEM map in Section 3.3. Vit-105 (Fig. 2c) shows a shift of the first peak to ~ 4 $nm^{-1}$ and much more subdued 2- and 6-fold intensities, indicating that the addition of extra elements, Ni and Ti, substantially frustrates topological MRO, especially the Cu-rich crystal-like MRO observed in $Zr_{50}Cu_{40}Al_{10}$.

As the composition shifts to $Zr_{60}Cu_{30}Al_{10}$ and $Zr_{65}Cu_{25}Al_{10}$ (Fig. 1d and 1e) with increasing Zr content, the position of the 2-fold peak gradually shifts to ~ 3.9 $nm^{-1}$, reflecting the Zr-rich nature of the local ordering. Both compositions lack clear 6-fold symmetry, indicating the absence of Cu-rich crystal-like MRO, as also observed in Vit-105. However, despite that, the 2-fold peak intensities of $Zr_{60}Cu_{30}Al_{10}$ and $Zr_{65}Cu_{25}Al_{10}$ are much higher than that of Vit-105 and comparable to that of $Zr_{50}Cu_{50}$ where the crystal-like MRO is prevalent. This indicates that the high 2-fold peak in these two compositions arises primarily from the geometric MRO measured in Fig. 1, rather than from topological MRO. Both alloys show longer $\zeta$ values, corresponding to a higher degree of geometric MRO. A longer $\zeta$ is expected to produce a higher first peak amplitude in the structure factor, $S(q)$ [10]. Since the angular correlation function is also measured in $q$ space (equivalently $k$ space in this work), the enhanced first peak in Fig. 2d and 2e likely reflects their higher degree of geometric MRO. This is also consistent with the high 1$^{st}$ peak amplitude that we observed in synchrotron $S(q)$ of $Zr_{65}Cu_{25}Al_{10}$ (Fig. S1).

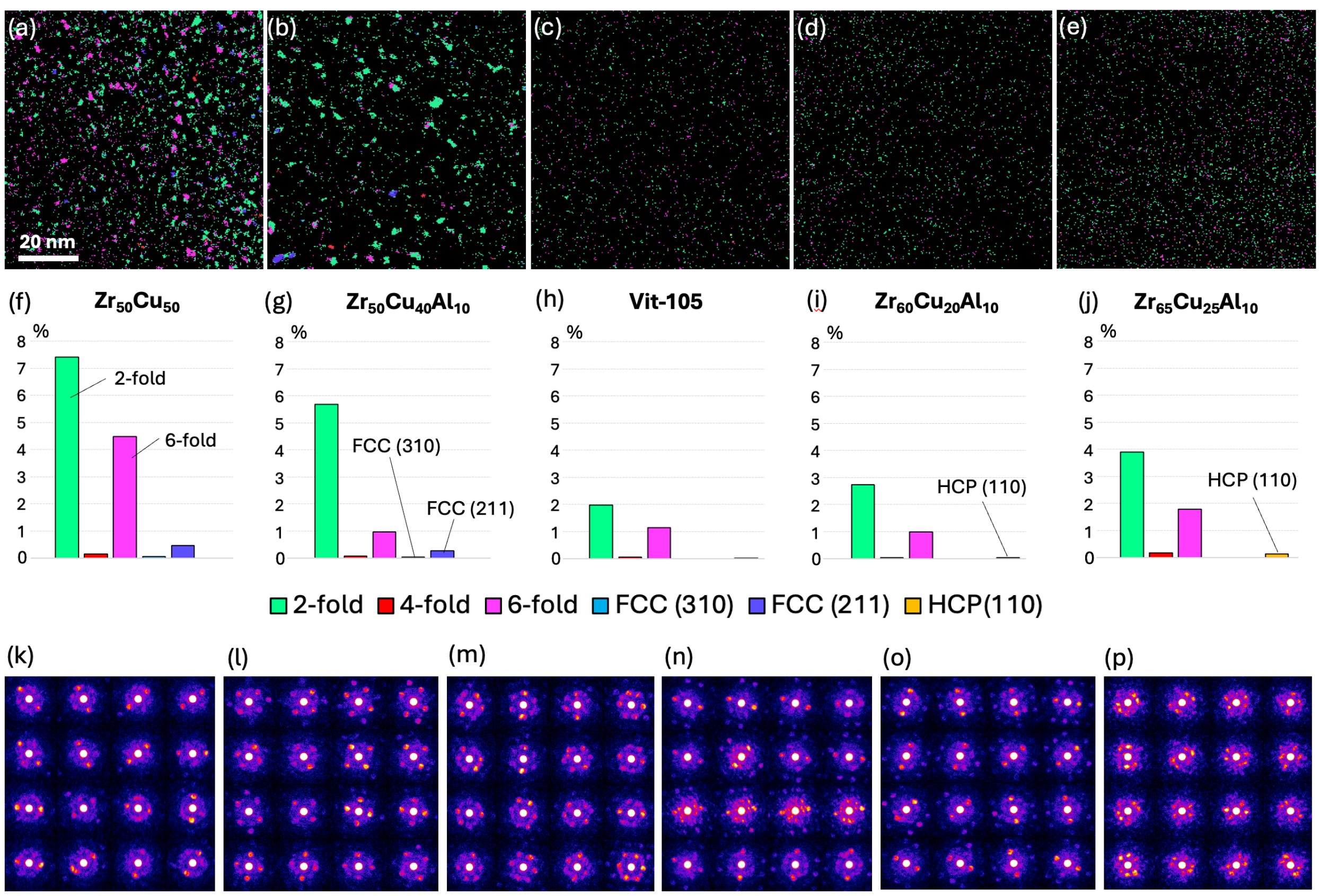

**Figure 3.** (a to e) Maps and (f to j) statistics of structural symmetries obtained from ML analysis of 4D-STEM. The color scale applies to both the maps and the statistics plots. Example nanodiffraction patterns showing (k) 2-fold, (l) 4-fold, (m) 6-fold, (n) FCC (310)-like, (o) FCC (211)-like, and (p) HCP (110)-like symmetries.

### 3.3. Structural symmetry maps and statistics from ML analysis

Figure 3 shows the spatial maps and statistics of symmetry types present in the 4D-STEM nanodiffraction patterns. In all five compositions, patterns with no symmetry, corresponding to the black background in the spatial maps, consistently account for the majority of patterns, as expected for disordered structures. However, a statistically meaningful fraction of symmetry-bearing patterns is observed, including 2-, 6-, and higher-order patterns. The relative fractions of these symmetries vary substantially across compositions.

The binary $Zr_{50}Cu_{50}$ (Fig. 3a) shows a high percentage of 2- and 6-fold patterns, consistent with the angular correlation features in Fig. 2a. It also exhibits a small number of FCC (310)-like and FCC (211)-like patterns (light blue and purple, respectively), as shown in the example patterns (Fig. 3n and 3o). These higher-order patterns directly confirm the presence of Cu-rich, FCC-like topological MRO domains at the nanoscale. In other words, all strong symmetries observed in $Zr_{50}Cu_{50}$, including 2-, 6-fold, and higher-order patterns, originate from the same Cu-rich FCC-like MRO, which dominates the nanoscale structure of this composition. This high fraction of crystal-like MRO may also be associated with the low glass-forming ability of $Zr_{50}Cu_{50}$, which has primarily been produced in thin ribbon form.

The ternary eutectic $Zr_{50}Cu_{40}Al_{10}$ (Fig. 3b) shows similar symmetry types to $Zr_{50}Cu_{50}$. Together with the identical peak position at 4.8 $nm^{-1}$ in the angular correlation function (Fig. 2b), this indicates that $Zr_{50}Cu_{40}Al_{10}$ is also rich in Cu-rich FCC-like MRO. However, the spatial and statistical distributions differ from those of $Zr_{50}Cu_{50}$. The fraction of all symmetry types, most notably for 6-fold symmetry, decreases in $Zr_{50}Cu_{40}Al_{10}$. In the spatial maps, 6-fold symmetry (magenta) is largely suppressed, while 2-fold regions appear larger than in $Zr_{50}Cu_{50}$. This may indicate that increased structural frustration due to the addition of 10% Al may disrupt 6-fold symmetry, causing the ML model to classify these regions as 2-fold rather than 6-fold. Within the density-wave framework for MGs [10], the large FCC-like MRO in $Zr_{50}Cu_{40}Al_{10}$ may not represent geometric MRO frozen in at the glass transition. Instead, it may reflect shorter-range topological or chemical ordering that contributes differently to the density correlations measured by $G(r)$. The atomic arrangements within these nanoscale regions may therefore be dominated by short-range bonding preferences that promote the local enrichment of certain atomic species, in this case primarily Cu atoms. While Ref. [10] emphasized ionic or covalent bonding as examples of short-range chemical bonding that can compete with density-wave MRO, our result shows that similar short-range chemical preferences can also arise from predominantly metallic bonding.

Vit-105 shows the lowest number of detected symmetries (Fig. 3c and 3h), with the 2-fold symmetry exhibiting the lowest percentage among all five compositions. The composition of Vit-105, $Zr_{52.5}Cu_{17.9}Ni_{14.6}Al_{10}Ti_5$, is similar to that of eutectic $Zr_{50}Cu_{40}Al_{10}$ in its Zr and Al contents, but with a portion of Cu replaced by Ni and Ti. The addition of these elements likely disrupts the large FCC-like MRO present in $Zr_{50}Cu_{40}Al_{10}$, resulting in a much more structurally frustrated nanostructure.

$Zr_{60}Cu_{30}Al_{10}$ also shows no large MRO domains in the spatial maps (Fig. 3d and 3e), leading to substantially fewer 2- and 6-fold patterns than in $Zr_{50}Cu_{40}Al_{10}$. This indicates that, as the composition becomes Zr-dominant, the formation of Cu-rich FCC-like MRO is significantly suppressed, which is also consistent with the disappearance of 6-fold symmetry in the angular correlation function in Fig. 2d. However, the fractions of 2- and 6-fold patterns in $Zr_{60}Cu_{30}Al_{10}$ are higher than those in Vit-105. This can be attributed to the small fraction of HCP(110)-like patterns detected in $Zr_{60}Cu_{30}Al_{10}$, shown by the yellow bar in the statistics plot in Fig. 3i and the example patterns in Fig. 3p. These results indicate the formation of small Zr-rich regions where the atoms exhibit chemical bonding that favors HCP-like packing, as in crystalline $\alpha$-Zr. The increase in 2- and 6-fold symmetries in $Zr_{60}Cu_{30}Al_{10}$ can therefore be understood as arising from these HCP-like packing motifs. Not surprisingly, with higher Zr content, the fraction of HCP-like patterns in $Zr_{65}Cu_{25}Al_{10}$ increases even further, which also increases the fractions of 2- and 6-fold patterns. (Fig 3e and 3j). This means that, with a mixture of more HCP-like packing, the nanoscale structure of $Zr_{65}Cu_{25}Al_{10}$ is more diversified in its local structural motifs than $Zr_{60}Cu_{30}Al_{10}$, which may have important implications for the mechanical properties of $Zr_{65}Cu_{25}Al_{10}$. This structure-property relationship will be explained in detail in Section 3.5.

The average ML classification confidence exceeded 90% for all symmetry classes in most compositions, with symmetry-containing classes consistently showing above 90% average confidence. These values indicate that, when symmetry is detected, the predictions are made with strong statistical certainty. However, $Zr_{60}Cu_{30}Al_{10}$ and Vit-105 show relatively lower confidence values, ranging between ~80 and 90%. We speculate that this may reflect increased structural frustration in these two compositions, which makes the diffraction patterns less distinguishable. A complete breakdown of class distributions and average confidence levels is provided in Table S1. The consistently high confidence scores across all symmetry classes and compositions suggest that the trained model reliably distinguishes meaningful patterns from background noise, *i.e.*, random scattering signals.

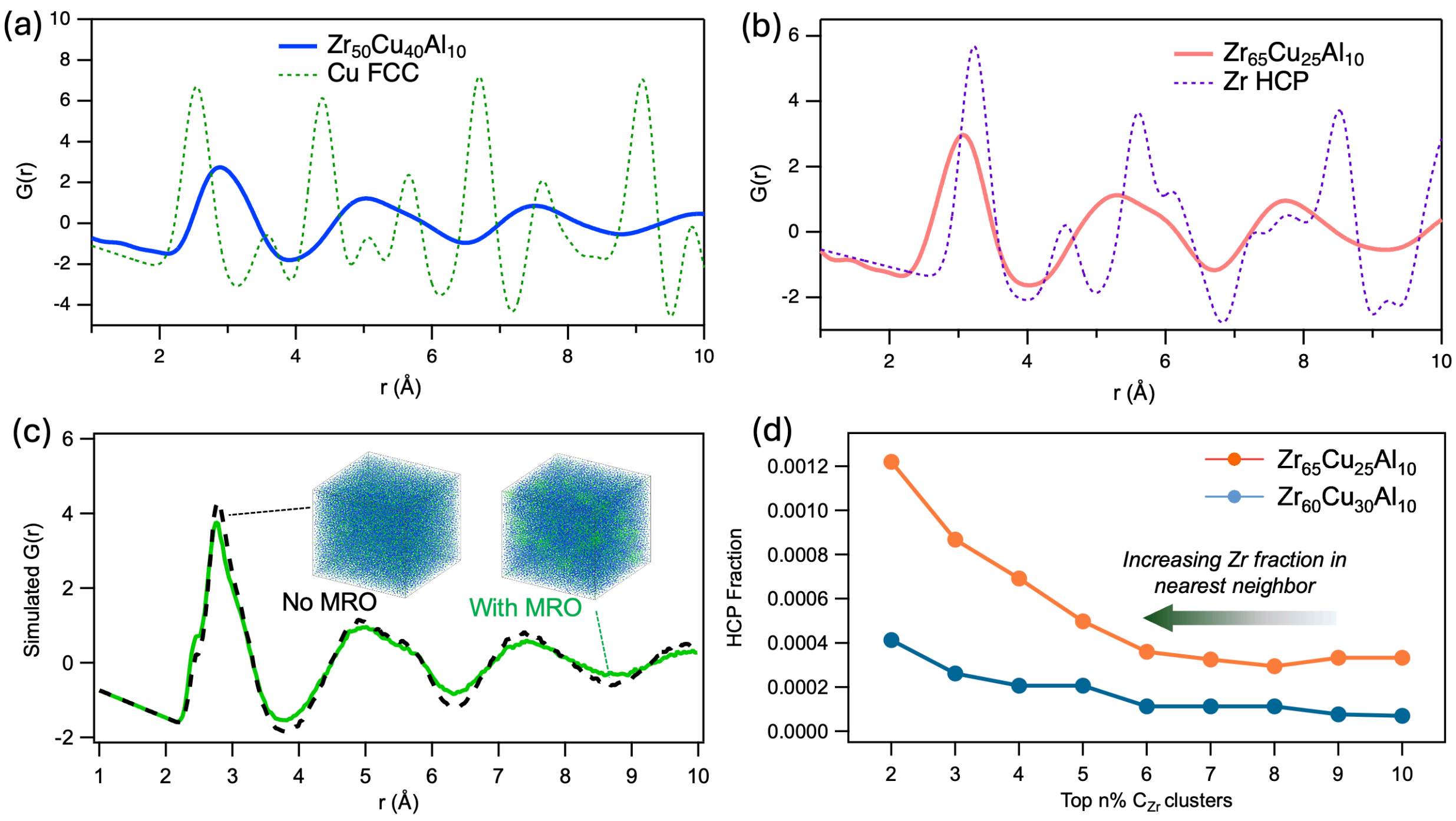

**Figure 4.** (a) Comparison between experimental $G(r)$ of $Zr_{50}Cu_{40}Al_{10}$ and simulated $G(r)$ of Cu FCC. (b) Same for $Zr_{65}Cu_{25}Al_{10}$ and Zr HCP. The simulated $G(r)$ from Cu FCC and Zr HCP were convoluted with a Gaussian broadening function with a width of 0.2 Å. (c) Simulated $G(r)$ from MG models with and without FCC-like MRO (inset). (d) Fraction of HCP-like common neighbor as a function of fraction of Zr in the nearest neighbor cluster ($C_{Zr}$), from simulated EAM-MD $Zr_{60}Cu_{30}Al_{10}$ and $Zr_{65}Cu_{25}Al_{10}$ models.

### 3.4. Effect of topological MRO on structural coherence length

Here, we explain how the presence of crystal-like MRO measured by 4D-STEM affects the overall coherence length, $\zeta$, measured by density wave analysis. Two representative compositions, $Zr_{50}Cu_{40}Al_{10}$ and $Zr_{65}Cu_{25}Al_{10}$, are used for this explanation. For $Zr_{50}Cu_{40}Al_{10}$, we argue that large Cu-rich, FCC-like topological MRO contributes to the short $\zeta$ measured, while such an effect is generally absent in $Zr_{65}Cu_{25}Al_{10}$, where HCP-like MRO is distributed with notably smaller size. This can be explained in part by comparing the measured $G(r)$ from each alloy to simulated $G(r)$ from Cu-FCC and Zr-HCP ordering, as shown in Fig. 4. The comparison between the $G(r)$ of $Zr_{50}Cu_{40}Al_{10}$ and that of its constituent FCC-like ordering (Fig. 4a) shows an overall qualitative mismatch, while the comparison between $Zr_{65}Cu_{25}Al_{10}$ and HCP ordering shows a much better match, especially in terms of the overall phase of the oscillation (Fig. 4b). This suggests that the large 1-5 nm FCC-like MRO in $Zr_{50}Cu_{40}Al_{10}$ may disrupt the overall structural coherence, which is not the case for the smaller 0.5-1 nm HCP-like ordering in $Zr_{65}Cu_{25}Al_{10}$. In other words, FCC-like MRO is structurally distinct from the surrounding glassy matrix of $Zr_{50}Cu_{40}Al_{10}$, reducing $\zeta$, while smaller HCP-like ordering is more integrated into the overall atomic environment and more compatible with the structure of $Zr_{65}Cu_{25}Al_{10}$, contributing to longer $\zeta$.

The shortening of $\zeta$ by large Cu-rich FCC-like MRO is also demonstrated using an atomistic model, as shown in Fig. 4c. An amorphous MG structure with $Zr_{50}Cu_{50}$ composition was generated using an EAM-potential-driven molecular dynamics (MD) model, containing 50,000 atoms within a 10 $nm^3$ volume (Fig. 4c inset, left). To mimic the crystal-like MRO observed in 4D-STEM, multiple 1 nm-diameter Cu-FCC crystals were then embedded in the model, followed by structural relaxation using MD (Fig. 4c inset, right). Both models retained the same composition. The corresponding $G(r)$ curves for these three models are shown in Fig. 4c. Notably, after relaxation with embedded MRO, the amplitude of $G(r)$ (green solid line) is lower than that of the original glassy $Zr_{50}Cu_{50}$ structure (black dashed line). This indicates that, within a given composition, incorporating crystal-like MRO into the glassy structure can reduce $\zeta$, as a shorter $\zeta$ corresponds to stronger damping of the $G(r)$ oscillations.

To confirm the effect of Zr content on the short-range chemical tendency toward HCP ordering, common neighbor analysis was performed using EAM models with 50,000 atoms to quantify the fraction of HCP-like ordering in $Zr_{60}Cu_{30}Al_{10}$ and $Zr_{65}Cu_{25}Al_{10}$ (Fig. 4d). As the fraction of Zr within nearest-neighbor bonding increases, $Zr_{65}Cu_{25}Al_{10}$ shows a significantly larger increase in HCP ordering than $Zr_{60}Cu_{30}Al_{10}$. The models demonstrate that higher Zr content leads to greater HCP-like motif formation, although the absolute value of the HCP fraction is smaller than the experimental counterpart (Fig. 3i and 3j), possibly due to the much faster quenching rate of the EAM MD simulation.

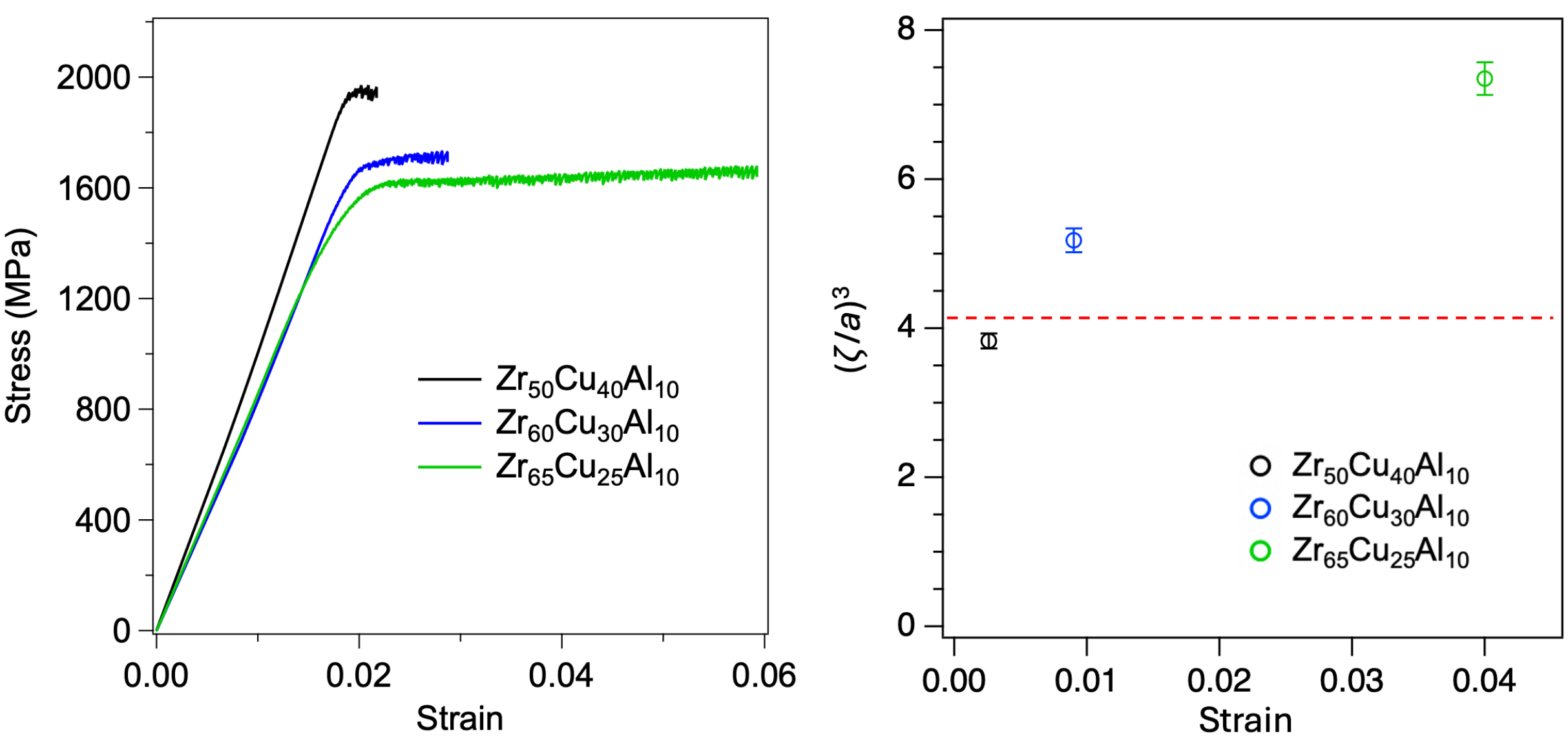

**Figure 5.** (a) Stress-strain curves of ZrCuAl MGs under study, adapted from Calderon Ortiz et al. [16]. (b) Coherence volume, $(\zeta/a)^3$ , vs. strain for the same MGs.

### 3.5 Discussion: MRO that governs ductility

Here, we correlate the measured MRO, including both geometric MRO from density wave analysis and topological MRO, with the mechanical behavior of the MGs under study. Figure 5 shows the stress-strain curves for $Zr_{50}Cu_{40}Al_{10}$, $Zr_{60}Cu_{30}Al_{10}$ and $Zr_{65}Cu_{25}Al_{10}$ measured in our previous work [16]. These three compositions show significantly different plastic strains of 0.026, 0.009, and 0.04, respectively. Mechanical data for Vit-105 were not measured in this work. Earlier reports showed highly scattered mechanical behavior for Vit-105, which was attributed to mechanisms beyond pure plastic deformation, such as cavity-induced deformation [34]. As shown in Fig. 1f, the coherence volumes $(\zeta/a)^3$ , for $Zr_{50}Cu_{40}Al_{10}$, Vit-105, $Zr_{60}Cu_{30}Al_{10}$ and $Zr_{65}Cu_{25}Al_{10}$ are $3.83 \pm 0.10$, $3.99 \pm 0.16$, $5.18 \pm 0.16$, and $7.35 \pm 0.22$, respectively.

We first compare the coherence volume and ductility of $Zr_{50}Cu_{40}Al_{10}$ and Vit-105. The two alloys have statistically indistinguishable $(\zeta/a)^3$ values of $3.83 \pm 0.10$ and $3.99 \pm 0.16$ within fitting uncertainty. These two values are only slightly below the $(\zeta/a)^3$ value of 4.2 (red dashed line in Fig. 5b) that has been suggested as the threshold between brittle and ductile tendencies [10]. However, the measured plastic strains are very different. $Zr_{50}Cu_{40}Al_{10}$ tends to be very brittle, with a plastic strain of only 0.026, while Vit-105 shows scattered but at least moderate plasticity [34]. This indicates that the very brittle behavior observed in $Zr_{50}Cu_{40}Al_{10}$ may be influenced by the large 1-5 nm Cu-rich, FCC-like topological MRO domains observed in this alloy (Figs. 3b and 3g), while such an effect is absent in Vit-105 because it lacks such large crystal-like MRO domains. This may explain the discrepancy in plasticity between these two compositions despite their similar coherence volumes. We speculate that the large crystal-like MRO domains tend to have stable internal bonding, which slows the dynamics of atomic rearrangement under load. However, this explanation may only be applicable to $Zr_{50}Cu_{40}Al_{10}$ where the crystal-like MRO domains are large in size, as we explain further below.

We then compare $Zr_{60}Cu_{30}Al_{10}$ and $Zr_{65}Cu_{25}Al_{10}$, both of which have coherence volumes above the $(\zeta/a)^3 \approx 4.2$ threshold. The coherence volume increases from $5.18 \pm 0.16$ for $Zr_{60}Cu_{30}Al_{10}$ to $7.35 \pm 0.22$ for $Zr_{65}Cu_{25}Al_{10}$, corresponding to an increase of approximately 40%. This increase is consistent with the higher ductility of $Zr_{65}Cu_{25}Al_{10}$, indicating that the density-wave coherence volume captures an important part of the ductility trend. However, the increase in plastic strain is much larger, from 0.009 for $Zr_{60}Cu_{30}Al_{10}$ to 0.04 for $Zr_{65}Cu_{25}Al_{10}$, corresponding to an increase of approximately 400%. Although density-wave analysis does not assume a quantitative linear relationship between coherence volume and plastic strain, our result suggests that the ductility increase from $Zr_{60}Cu_{30}Al_{10}$ to $Zr_{65}Cu_{25}Al_{10}$ may involve structural factors beyond the increase in density-wave coherence volume. We argue that the enhanced ductility of $Zr_{65}Cu_{25}Al_{10}$ is likely aided by its higher fraction of small, Zr-rich HCP-like topological MRO domains (Fig. 3e and 3j). Unlike the large Cu-rich FCC-like MRO domains in $Zr_{50}Cu_{40}Al_{10}$, these HCP-like domains are much smaller, on the order of ~1 nm or less, and are therefore expected to diversify the local structure without forming large rigid regions. Using the diffraction criteria we explained in Im et al. [14], which involves a 6% diffraction sampling limit and the given geometry of the probed volume ($80 \times 80 \times 25$ nm$^3$), we estimate that these Zr-rich HCP-like topological MRO domains constitute about 5% of the total volume of $Zr_{65}Cu_{25}Al_{10}$, which is high enough to produce meaningful diversification of the nanostructure compared with $Zr_{60}Cu_{30}Al_{10}$.

Such nanoscale diversification may promote ductility through two related mechanisms. First, as suggested by mesoscale simulations incorporating diversified MRO structures [35, 36], structural heterogeneity at the nanoscale can frustrate strain propagation between shear transformation zones (STZs). In those simulations, motivated by the presence of crystal-like MRO [23], the glass structure was modeled as a distribution of STZs with different activation energies and softening behaviors. This produced more spatially dispersed shear-band patterns, indicating enhanced ductility without rapid localized failure. The fine-scale structural diversification in $Zr_{65}Cu_{25}Al_{10}$ may therefore promote more distributed plastic deformation, contributing to the significant difference between $Zr_{60}Cu_{30}Al_{10}$ and $Zr_{65}Cu_{25}Al_{10}$.

Second, the effect of HCP-like MRO can also be considered from the viewpoint of the evolving structure within shear bands. While the initial nanoscale structure may influence shear-band initiation through strain frustration, subsequent plastic deformation is governed by repeated reactivation of shear bands during serrated flow. In this regime, the structural state within the shear band becomes critical. Prior works have shown that shear-band dynamics in this system are thermally activated and strongly composition dependent, with increasing Zr content leading to higher activation barriers and slower shear-band propagation [37, 38]. This suggests that shear propagation becomes more resistant with increasing Zr content. At the same time, Zr-rich compositions promote HCP-like local ordering, which may provide a stronger driving force for structural relaxation within the shear band during deformation. Thus, although shear may become more difficult to initiate or propagate, the material inside the activated shear band may relax more efficiently, enabling repeated shear events and higher accumulated plastic strain. From this viewpoint, $Zr_{65}Cu_{25}Al_{10}$ represents the composition in which a larger geometric coherence volume and fine-scale, chemically compatible HCP-like topological MRO act in the same direction. The former increases the cooperative length scale for ductile relaxation, while the latter helps distribute and repeatedly reactivate shear-band deformation.

**5. Conclusion**

In this work, we show that the composition-dependent ductility of Zr-based metallic glasses is governed by a coupled mechanism between geometric and topological MRO, rather than by either structural descriptor alone. By combining density-wave analysis of synchrotron $G(r)$ with ML-assisted 4D-STEM, we distinguish geometric MRO, represented by the density-correlation coherence volume, from topological MRO, represented by nanoscale symmetry-bearing local motifs. The density-wave coherence volume captures an important baseline tendency for ductile relaxation.

However, comparisons between compositions with similar coherence volumes, such as $Zr_{50}Cu_{40}Al_{10}$ and Vit-105, and between compositions with different topological MRO, such as $Zr_{60}Cu_{30}Al_{10}$ and $Zr_{65}Cu_{25}Al_{10}$, show that local motif topology substantially influences how this geometric coherence is expressed mechanically. Large Cu-rich FCC-like MRO domains in $Zr_{50}Cu_{40}Al_{10}$ appear to act as structurally distinct, relatively rigid regions that reduce structural coherence and promote brittle behavior, whereas fine, Zr-rich HCP-like MRO in $Zr_{65}Cu_{25}Al_{10}$ provides nanoscale structural diversification that can frustrate strain localization and facilitate repeated shear-band relaxation. These results establish a structure-property relationship in which metallic-glass ductility is controlled by both the cooperative length scale of density fluctuations and the size, chemistry, and distribution of local topological MRO motifs.

**Acknowledgement**

The authors acknowledge the support of NSF-DMR-2406530 (M.I., Yunzhi W., J. H.) and NSF-DMR-2406531 (Muchen W., Y. F.). G.Y., J.K., and E.P were supported by the Creative Materials Discovery Program through the National Research Foundation of Korea (NRF) funded by the Ministry of Science and ICT (No. NRF-2019M3D1A1079215), Samsung Research Funding Center of Samsung Electronics under SRFC-MA1802–06. H.C. and S.L. were supported by the Basic Science Research Program (C624100) funded by the Korea Basic Science Institute. Electron microscopy was performed in the Center for Electron Microscopy and Analysis at The Ohio State University.

**References**

1. B. Zhang, D.Q. Zhao, M.X. Pan, W.H. Wang, and A.L. Greer, *Amorphous Metallic Plastic*. Physical Review Letters, 2005. **94**(20): p. 205502.
2. Y.H. Liu, G. Wang, R.J. Wang, D.Q. Zhao, M.X. Pan, and W.H. Wang, *Super Plastic Bulk Metallic Glasses at Room Temperature*. Science, 2007. **315**(5817): p. 1385-1388.
3. J.M. Park, J.H. Han, N. Mattern, D.H. Kim, and J. Eckert, *Designing Zr-Cu-Co-Al Bulk Metallic Glasses with Phase Separation Mediated Plasticity*. Metallurgical and Materials Transactions A, 2012. **43**(8): p. 2598-2603.
4. Y. Fan, T. Iwashita, and T. Egami, *Energy landscape-driven non-equilibrium evolution of inherent structure in disordered material*. Nature Communications, 2017. **8**: p. 15417.
5. J. Ding, S. Patinet, M.L. Falk, Y. Cheng, and E. Ma, *Soft spots and their structural signature in a metallic glass*. Proceedings of the National Academy of Sciences, 2014. **111**(39): p. 14052.
6. M. Wang, Y. Wang, M. Islam, Y. Wang, Y. Wang, J. Hwang, and Y. Fan, *Dual machine learning pinpoints the Radius of Informative Structural Environments in metallic glasses*. npj Computational Materials, 2026.
7. C. Liu, Y. Wang, Y. Wang, M. Islam, J. Hwang, Y. Wang, and Y. Fan, *Concurrent prediction of metallic glasses' global energy and internal structural heterogeneity by interpretable machine learning*. Acta Materialia, 2023. **259**: p. 119281.
8. *Chapter 3 The method of total scattering and atomic pair distribution function analysis*, in *Pergamon Materials Series*, T. Egami and S.J.L. Billinge, Editors. 2003, Pergamon. p. 55-99.
9. R. Dai, A.K. Gangopadhyay, R.J. Chang, and K.F. Kelton, *A method to predict the glass transition temperature in metallic glasses from properties of the equilibrium liquid*. Acta Materialia, 2019. **172**: p. 1-5.
10. T. Egami, W. Dmowski, and C.W. Ryu *Medium-Range Order Resists Deformation in Metallic Liquids and Glasses*. Metals, 2023. **13**, DOI: 10.3390/met13030442.
11. T. Egami and C.W. Ryu, *World beyond the nearest neighbors*. Journal of Physics: Condensed Matter, 2023. **35**(17): p. 174002.
12. C.W. Ryu, W. Dmowski, K.F. Kelton, G.W. Lee, E.S. Park, J.R. Morris, and T. Egami, *Curie-Weiss behavior of liquid structure and ideal glass state*. Scientific Reports, 2019. **9**(1): p. 18579.
13. C.W. Ryu and T. Egami, *Alpha-relaxation by scattering and medium-range atomic correlation in simple liquids*. The Journal of Chemical Physics, 2025. **163**(14): p. 144506.
14. S. Im, Z. Chen, J.M. Johnson, P. Zhao, G.H. Yoo, E.S. Park, Y. Wang, D.A. Muller, and J. Hwang, *Direct determination of structural heterogeneity in metallic glasses using four-dimensional scanning transmission electron microscopy*. Ultramicroscopy, 2018. **195**: p. 189-193.
15. S. Im, Y. Wang, P. Zhao, G.H. Yoo, Z. Chen, G. Calderon, M. Abbasi Gharacheh, M. Zhu, O. Licata, B. Mazumder, D.A. Muller, E.S. Park, Y. Wang, and J. Hwang,

*Medium-range ordering, structural heterogeneity, and their influence on properties of Zr-Cu-Co-Al metallic glasses.* Physical Review Materials, 2021. **5**(11): p. 115604.
16. G.A.C. Ortiz, M. Islam, G.H. Yoo, J.Y. Kim, S. Im, Y. Wang, Y. Wang, Y. Fan, Y. Wang, E.S. Park, and J. Hwang, *Substantial change in medium range ordering and its influence on glass forming ability and mechanical properties of ZrCu and ZrCuAl metallic glasses.* Acta Materialia, 2025. **298**: p. 121402.
17. B. Riechers, T.C. Pekin, X.-Y. Luo, Y. Sun, C.T. Koch, P.M. Derlet, and R. Maaß, *Spatial distribution and connectivity of medium-range order signatures in a metallic glass probed with simulated and experimental 4DSTEM.* Journal of Alloys and Compounds, 2026. **1058**: p. 186631.
18. S. Kang, V. Wollersen, C. Minnert, K. Durst, H.-S. Kim, C. Kübel, and X. Mu, *Mapping local atomic structure of metallic glasses using machine learning aided 4D-STEM.* Acta Materialia, 2024. **263**: p. 119495.
19. K. Nakazawa, K. Mitsuishi, K. Iakoubovskii, S. Kohara, and K. Tsuchiya, *Structure-dynamics relation in metallic glass revealed by 5-dimensional scanning transmission electron microscopy.* NPG Asia Materials, 2024. **16**(1): p. 57.
20. S. Huang and P.M. Voyles, *Momentum transfer resolved electron correlation microscopy.* Ultramicroscopy, 2024. **256**: p. 113886.
21. C. Francis and P.M. Voyles, *Clustering characteristic diffraction vectors in 4-D STEM data sets from overlapping structures in nanocrystalline and amorphous materials.* Ultramicroscopy, 2024. **267**: p. 114040.
22. P. Voyles and J. Hwang, *Fluctuation Electron Microscopy*, in *Characterization of Materials*. 2012. p. 1-7.
23. J. Hwang, Z.H. Melgarejo, Y.E. Kalay, I. Kalay, M.J. Kramer, D.S. Stone, and P.M. Voyles, *Nanoscale Structure and Structural Relaxation in Zr50Cu45Al5 Bulk Metallic Glass.* Physical Review Letters, 2012. **108**(19): p. 195505.
24. V. Schmidt, H. Rösner, M. Peterlechner, G. Wilde, and P.M. Voyles, *Quantitative Measurement of Density in a Shear Band of Metallic Glass Monitored Along its Propagation Direction.* Physical Review Letters, 2015. **115**(3): p. 035501.
25. J. Hwang and P.M. Voyles, *Variable Resolution Fluctuation Electron Microscopy on Cu-Zr Metallic Glass Using a Wide Range of Coherent STEM Probe Size.* Microscopy and Microanalysis, 2011. **17**(1): p. 67-74.
26. J. Hwang, A.M. Clausen, H. Cao, and P.M. Voyles, *Reverse Monte Carlo structural model for a zirconium-based metallic glass incorporating fluctuation microscopy medium-range order data.* Journal of Materials Research, 2009. **24**(10): p. 3121-3129.
27. P.M. Voyles, J.L. Grazul, and D.A. Muller, *Imaging individual atoms inside crystals with ADF-STEM.* Ultramicroscopy, 2003. **96**(3): p. 251-273.
28. G.A. Calderón Ortiz, M. Zhu, A. Wadsworth, L. Dou, I. McCulloch, and J. Hwang, *Unveiling Nanoscale Ordering in Amorphous Semiconducting Polymers Using Four-Dimensional Scanning Transmission Electron Microscopy.* ACS Applied Materials & Interfaces, 2024. **16**(41): p. 55852-55863.

29. M. Abbasi, Y. Dong, J. Meng, D. Morgan, X. Wang, and J. Hwang, *In situ observation of medium range ordering and crystallization of amorphous TiO2 ultrathin films grown by atomic layer deposition.* APL Materials, 2023. **11**(1): p. 011102.
30. H. Tanaka, T. Kawasaki, H. Shintani, and K. Watanabe, *Critical-like behaviour of glass-forming liquids.* Nature Materials, 2010. **9**(4): p. 324-331.
31. M. Zhu, J. Lanier, J. Flores, V. da Cruz Pinha Barbosa, D. Russell, B. Haight, P.M. Woodward, F. Yang, and J. Hwang, *Structural degeneracy and formation of crystallographic domains in epitaxial LaFeO3 films revealed by machine-learning assisted 4D-STEM.* Scientific Reports, 2024. **14**(1): p. 4198.
32. D.T. Schweiss, J. Hwang, and P.M. Voyles, *Inelastic and elastic mean free paths from FIB samples of metallic glasses.* Ultramicroscopy, 2013. **124**: p. 6-12.
33. S.C. Glade, R. Busch, D.S. Lee, W.L. Johnson, R.K. Wunderlich, and H.J. Fecht, *Thermodynamics of Cu47Ti34Zr11Ni8, Zr52.5Cu17.9Ni14.6Al10Ti5 and Zr57Cu15.4Ni12.6Al10Nb5 bulk metallic glass forming alloys.* Journal of Applied Physics, 2000. **87**(10): p. 7242-7248.
34. A. Das, C. Ott, D. Pechimuthu, R. Moosavi, M. Stoica, P.M. Derlet, and R. Maaß, *Shear-band cavitation determines the shape of the stress-strain curve of metallic glasses.* Physical Review Materials, 2023. **7**(2): p. 023602.
35. P. Zhao, J. Li, J. Hwang, and Y. Wang, *Influence of nanoscale structural heterogeneity on shear banding in metallic glasses.* Acta Materialia, 2017. **134**: p. 104-115.
36. Y. Wang, Y. Wang, C. Liu, J. Hwang, Y. Fan, and Y. Wang, *Atomistically informed mesoscale modelling of deformation behavior of bulk metallic glasses.* Acta Materialia, 2024. **276**: p. 120136.
37. P. Thurnheer, R. Maaß, S. Pogatscher, and J.F. Löffler, *Compositional dependence of shear-band dynamics in the Zr–Cu–Al bulk metallic glass system.* Applied Physics Letters, 2014. **104**(10): p. 101910.
38. P. Thurnheer, R. Maaß, K.J. Laws, S. Pogatscher, and J.F. Löffler, *Dynamic properties of major shear bands in Zr–Cu–Al bulk metallic glasses.* Acta Materialia, 2015. **96**: p. 428-436.